\documentclass[prl,twocolumn,letterpaper,showpacs,superscriptaddress]{revtex4}
\usepackage{graphicx,bm,amssymb,amsmath}
\usepackage[dvips,colorlinks,linkcolor=black,citecolor=black,urlcolor=black]{hyperref}

\newcommand{\dd}{\text{d}}

\newcommand{\h}{\frac{1}{2}}

\def\be{\begin{equation}}
\def\fin{\end{equation}}

\def\T{{\sf T\kern-.45em T}}
\def\C{\kern.1em{\raise.47ex\hbox{$\scriptscriptstyle |$}}
\kern-.40em{\sf C}}

\begin{document} \title{Rebuilding cytoskeleton roads: Active-transport-induced polarization of cells} \date{\today} 
\author{ R.J. Hawkins}
\affiliation{UMR 7600, Universit\'e Pierre et Marie Curie/CNRS, 4 Place Jussieu, 75255
Paris Cedex 05 France}
\author{ O. B\'enichou}
\affiliation{UMR 7600, Universit\'e Pierre et Marie Curie/CNRS, 4 Place Jussieu, 75255
Paris Cedex 05 France}
\author{M. Piel}
\affiliation{UMR 144, Institut Curie/CNRS, 
26 rue d'Ulm 75248 Paris Cedex 05 France}
\author{ R.Voituriez}
\affiliation{UMR 7600, Universit\'e Pierre et Marie Curie/CNRS, 4 Place Jussieu, 75255
Paris Cedex 05 France}

\begin{abstract}
Many cellular processes require a polarization axis which generally initially emerges as an inhomogeneous distribution of molecular markers in the cell. We present a simple analytical model of a general mechanism of cell polarization taking into account the positive feedback due to the coupled dynamics of molecular markers and cytoskeleton filaments. 
We find that the geometry of the organization of  cytoskeleton filaments, nucleated on the membrane (e.g. cortical actin) or from a center in the cytoplasm (e.g. microtubule asters), dictates whether the system is capable of spontaneous polarization or polarizes only in response to external asymmetric signals. Our model also captures the main features of recent experiments of cell polarization in two considerably different biological systems; namely mating budding yeast and neuron growth cones.
\end{abstract}
\maketitle

Cell polarization is an essential step for many biological processes such as cell migration, division, development and morphogenesis in widely varying cell types. Polarization  is characterized in its early stages by an inhomogeneous distribution of specific molecular markers. Often cell polarization is driven by an external asymmetric signal as in the well known example of cells migrating along a chemical gradient performing chemotaxis \cite{alberts}. In mating yeast the external signal is a pheromone gradient, which causes the cell to grow an elongation known as a shmoo in the direction of the pheromone source \cite{alberts}. In the case of nerve cells, external gradients direct the growth of the growth cone \cite{Bouzigues2007}.  However observations show that some systems can also polarize spontaneously in the absence of any external gradient, e.g. as reported in neutrophil cell migration, algae cell division, 
or mating yeast \cite{Wedlich-Soldner2003a,Altschuler2008,Marco2007,Wedlich-Soldner2003}.
These two distinct polarization processes, {\it driven} or {\it spontaneous}, are necessary for  cells to fulfill different biological functions, but 
what determines whether a cell can polarize spontaneously or only in response to an external signal is still not well understood. In this Rapid Communication we propose a general physical mechanism to address this fundamental biological question, briefly discussing budding mating yeast and neuron growth cones as specific examples.

The molecular basis of spontaneous and driven cell polarization has been much debated over the past decade, and is likely to involve several pathways including historical marks (such as that left by the previous budding event in yeast \cite{alberts}). 
However, it is widely recognized that the cytoskeleton plays a crucial role in cell polarization. The efficiency of formation of polar caps in yeast is reduced when actin transport is disrupted 
and the polar caps formed are unstable \cite{Wedlich-Soldner2003,Wedlich-Soldner2004,Irazoqui2005}. In the case of neurons, it has been shown that the  polarization of the growth cone is suppressed when microtubules are depolymerized 
\cite{Bouzigues2007}.
To account for these observations, it is generally argued that the cytoskeleton filaments mediate an effective positive feedback in the dynamics of polarization markers \cite{Marco2007,Wedlich-Soldner2003,Piel2008}. 
This arises from the molecular markers not only diffusing in the cell cytoplasm, but also being actively transported by molecular motors along cytoskeleton filaments, the dynamic organization of which is regulated by the markers themselves (see \cite{Doubrovinski2007} for another example).

A number of recent studies have embarked upon modeling aspects of polarization.
Many are reaction-diffusion systems in which polarization emerges as a type of Turing instability \cite{Iglesias2008,Ma2004,Levine2006,Xu2003,Narang2006,Onsum2007} or due to stochastic fluctuations \cite{Altschuler2008} and some \cite{Onsum2007,Marco2007,Wedlich-Soldner2003} include cytoskeleton proteins as a regulatory factor. In particular \citet{Wedlich-Soldner2003} consider a one-dimensional (1D) model which shows that a positive feedback favors the emergence of polarized states, 
 however, as in other existing models,  the dynamics of markers in  the cytoplasm and the geometry of the filaments are not considered.
Here we propose an analytical model which explicitly addresses the geometry of the cytoskeleton and its dynamical coupling to the transport of markers while remaining general enough to be applicable to a wide range of cell systems.
Our model provides a minimal mechanism of cell polarization induced by active transport and  shows that the dynamical organization of cytoskeletal filaments actually plays a crucial role, since it dictates the polarization ability of cells -- spontaneous or driven. 
Specifically, our analysis
suggests that in general, the case of microtubule mediated transport does not lead to spontaneous polarization but the response to external signals can be accurate; while the case of cortical actin mediated transport leads to spontaneous polarization but a less robust response to gradients.

\begin{figure}[hbt]
\begin{centering}
\includegraphics[width=0.45\textwidth]{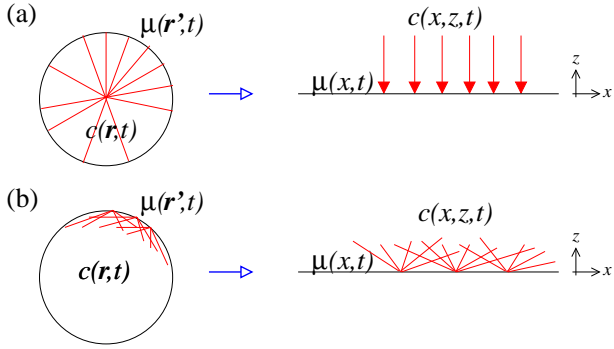}
\caption{\label{fig:geom}Model geometry.  Filaments nucleated at the (a) cell center and (b) membrane. Arrows indicate the velocity of directed transport. Concentration of molecular markers $\mu$ on the membrane and $c$ in the bulk.}
\end{centering}
\end{figure}

We model polarization markers as particles which can be
either on the membrane or in the cytoplasm of the cell.  For
simplicity  we assume that the cell is essentially bi-dimensional and we  neglect curvature effects. The membrane boundary is then taken as  a 1D line along the $x$-axis and the  cytoplasm  is parametrized by $(x,z)$. This geometry is sketched
in Fig.~\ref{fig:geom}.
The dynamical equations for the concentration of markers on the
membrane, $\mu(x,t)$, and in the  cytoplasm, $c(x,z,t)$,
are given by;
\begin{align}
\partial_t\mu(x,t)=&D_m\partial_x^2\mu(x,t)+j_m\label{eq:mudyn}\\
\partial_tc(x,z,t)=&-{\bf \nabla\cdot j_b}\label{eq:cdyn}
\end{align}
where $j_{m,b}$ are the flux of molecules arriving at the membrane and the flux in the bulk respectively;
\begin{align} 
j_m=&k^{\text{on}}c(x,0,t)-k^{\text{off}}\mu(x,t)\\ 
{\bf j_b}=&-D_b{\bf \nabla} c(x,z,t)+{\bf v}c(x,z,t).
\end{align} 
$D_{m,b}$ are the diffusion constants in the membrane
and the bulk of the cell respectively and $k^{\text{on,off}}$ are the
rates of attachment and detachment to the membrane. In the bulk we assume that
the rate of molecules switching between filament bound and free is high such that the flux of particles in the cell bulk, ${\bf
j_b}$, is effectively directed diffusion where ${\bf v}$ is the velocity of
directed transport of markers along cytoskeleton filaments \cite{Nedelec2001,Loverdo2008}. 

The directed transport ${\bf v}$, which drives the system out of equilibrium,  depends on the  geometry of filaments, which we assume to be determined by the concentration of markers on the membrane $\mu(x,t)$. This  makes the term involving ${\bf v}$ in Eq.~\ref{eq:cdyn} a nonlinear coupling term. We consider two types of idealized geometries for active transport motivated by the two general biological classes of (a) microtubule cytoskeleton and (b) cortical actin systems.

(a) Filaments that grow from a nucleating center  in the cytoplasm towards the membrane can be taken in a first approximation to be perpendicular to the membrane as sketched in Fig.~\ref{fig:geom}a. Assuming a ``search and capture model'' \cite{Mimori-Kiyosue2003}, their orientation is regulated by the polarity markers due to preferential stabilization and growth in regions of high marker concentration. With this geometry, in our model, the velocity of directed transport can be written simply as ${\bf
v}=-\alpha\mu(x,t){\bf{u}}_z$ where $\alpha$ is a parameter for the strength of coupling. 
In general microtubules are found with such geometry, e.g. in neuron growth cones, for which observations suggest that the polarity markers, GABA receptors, are associated with and regulate the growing microtubule ends 
\cite{Bouzigues2007,Dent2003}.

(b) Filaments that are polymerized from nucleators localized at the membrane can be approximated by a superposition of  asters centered on the
membrane as sketched in Fig.~\ref{fig:geom}b. We model this as a field at position ${\bf{r}}=(x,z)$
proportional to the filament concentration which decreases with distance from the nucleation point
${\bf{r^{\prime}}}=(x^{\prime},z^{\prime})$ on the membrane such that:
\begin{equation} 
{\bf v}=-\alpha\int_{-L/2}^{L/2}\frac{{\bf{r}}-{\bf{r^{\prime}}}}{|{\bf{r}}-{\bf{r^{\prime}}}|^2}\mu(x^{\prime},t)\dd
x^{\prime}\label{eq:v}
\end{equation}
where $\alpha$ is  the coupling parameter and $L$ is the cell perimeter. Note that we expect that other decaying functions of $|{\bf{r}}-{\bf{r^{\prime}}}|$ 
would not qualitatively change the results. 
This geometry mimics the organization of cortical actin (see  also \cite{Salbreux2007}), for instance in budding yeast \cite{Wedlich-Soldner2003,Karpova1998} in which molecular markers (e.g. Cdc42, Spa2, septins) are transported along the filaments towards the membrane by myosin V molecular motors. When active, these molecules in turn induce actin nucleation at the membrane thereby creating a positive feedback.

Finally, the above equations are completed by the condition for the
conservation of total number of polarity markers:
$M=\int_{\scriptscriptstyle{-L/2}}^{\scriptscriptstyle{L/2}}\dd x \,\mu(x,t) +\int_{\scriptscriptstyle{-L/2}}^{\scriptscriptstyle{L/2}}\dd x\int_{\scriptscriptstyle{0}}^{\scriptscriptstyle{\infty}}\dd z  \,  c(x,z,t)$. 
Note that we can assume that the system is not bounded in the $z$ direction since we will only consider concentration profiles exponentially decaying with $z$.

\paragraph{Spontaneous polarization.} To determine whether spontaneous polarization occurs in such systems we perform a linear stability analysis  of
Eqs. (\ref{eq:mudyn}-\ref{eq:cdyn}) around the  homogeneous out of equilibrium steady state solution which reads 
$\mu^0=\frac{k^{\text{on}}}{k^{\text{off}}}c^0(0)$, 
$c^0(z)=c^0(0)e^{-\lambda z}.$ 
Here $\lambda=\alpha\mu^0/D_b$ for case (a) and $\lambda=\alpha\mu^0\pi/D_b$ for case (b).  
The conservation of 
markers  fixes the value of $c^0(0)$ in
terms  $M$. We perturb the system in the infinite size $L$ limit about the
homogeneous steady state 
such that
$\mu(x,t)=\mu^0+\mu_ke^{ikx+st}$,  
$c(x,z,t)=c^0(z)+c_k(z)e^{ikx+st}$, 
and obtain to linear order from (\ref{eq:mudyn}):
\begin{equation}
\mu_k=\frac{k^{\text{on}}c_k(0)}{s+D_mk^2+k^{\text{off}}}\label{eq:muk}.
\end{equation}

In case (a) of a schematic  microtubule system where  ${\bf v}=-\alpha\mu(x,t){\bf{u}}_z$,  
Eq. (\ref{eq:cdyn}) 
is solved by $c_k(z)=Ae^{-\lambda z}+(c_k(0)-A)e^{-\rho z}$ with $\rho=\frac{\lambda}{2}\pm\h
(\lambda^2+4k^2+4s/D_b)^{1/2}$ and where $A$ is  determined by the 
conservation of flux at the membrane $(j_m+{\bf j_b}\cdot
{\bf{u}}_z)|_{z=0}=0$. Substituting this into (\ref{eq:cdyn}) leads to the dispersion relation for $s(k)$;
\begin{align}
(s+D_b k^2)&\left (k^{\text{off}}(2\lambda-\rho)+(s+D_m k^2)(\lambda-\rho-\frac{k^{\text{on}}}{D_b})\right )\nonumber\\
&+\lambda^2D_bk^{\text{off}}(\lambda-\rho)=0.\label{eq:s(k)perp}
\end{align}
In the limits of both small and large $k$ this gives the same solution $s=-Dk^2$ where $D=\text{min}(D_b,D_m)$. These limiting negative solutions suggest that the real part of $s(k)$ is always negative, which can indeed be checked by a numerical search of the parameter space. This shows that  the homogeneous state 
is linearly stable and there is no spontaneous
polarization. This actually compares well to experiments in neuron growth cones \cite{Bouzigues2007}.
When the gradient was removed, the distribution of receptors returned to a symmetric distribution indicating the stability of the homogeneous state and the lack of spontaneous polarization, as expected from our results.

\begin{figure}[hbt]
\begin{centering}
\includegraphics[width=0.5\textwidth]{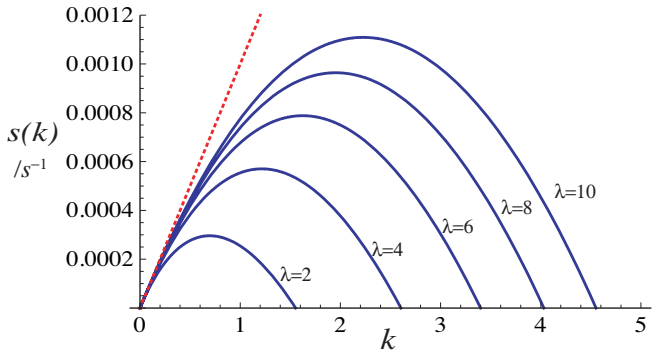}
\caption{\label{fig:s-k} Solution $s(k)$ of Eq.~(\ref{eq:s(k)}). $k$ and $\lambda$ are plotted in dimensionless units normalized by the inverse of the cell radius, $R^{-1}=2\pi/L$. Parameter values used are: $k^{\text{on}}=1\mu{\text m}{\text s}^{-1}$, $k^{\text{off}}=0.1{\text s}^{-1}$, $D_b=0.1\mu{\text m}^2{\text s}^{-1}$, $D_m=0.01\mu{\text m}^2{\text s}^{-1}$ and $R=10\mu{\text m}$. The dotted line shows the small $k$ asymptote $s=\frac{D_b k^{\text{off}}}{k^{\text{on}}}k$. }
\end{centering}
\end{figure}

\begin{figure}[hbt]
\begin{centering}
\includegraphics[width=0.4\textwidth]{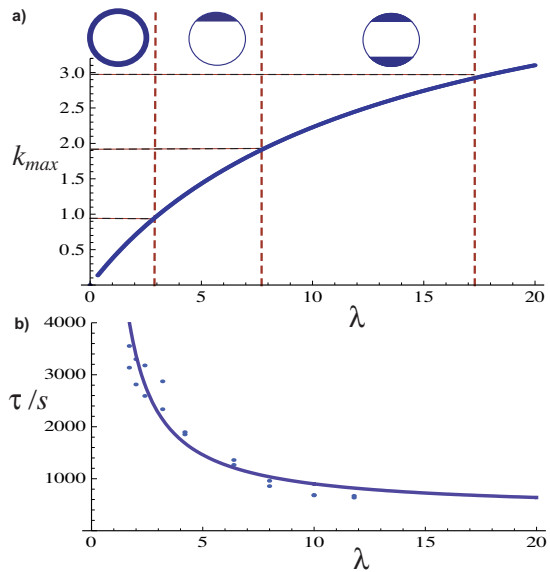}
\caption{\label{fig:kmax-lambda} a) The most unstable mode $k_{\text{max}}$ as a function of $\lambda=\alpha\mu^0\pi/D_b$ (both normalized by $R^{^-1}$ as in Fig.~\ref{fig:s-k}). Cartoons of the distribution of markers are shown above the curve. b) $\tau=1/s(k_{\text{max}})$ as a function $\lambda$. The calculated solid line is compared to data (points) from \cite{Piel2008} for the measured polarization time $t$ against pheromone concentration, $c_{\text{ph}}$. We used a scaling of $\lambda=2c_{\text{ph}}$ and $\tau=0.14t$ to obtain the observed fit. Other parameter values used are as in Fig.~\ref{fig:s-k}.}
\end{centering}
\end{figure}

We now consider case (b) of a schematic cortical actin system 
modeled by Eq.~(\ref{eq:v}). In this case Eq. (\ref{eq:cdyn}) 
is solved by $c_k(z)=Be^{-(\lambda
+k)z}+(c_k(0)-B)e^{-\rho z}$,  where $B$ is determined by the conservation of flux at the membrane 
and the dispersion relation finally reads:
\begin{align}
(s-\lambda D_b k)&\left (k^{\text{off}}(2\lambda-\rho)+(s+D_m k^2)(\lambda-\rho-\frac{k^{\text{on}}}{D_b})\right )\nonumber\\
&+\lambda^2D_bk^{\text{off}}(\lambda+k-\rho)=0.\label{eq:s(k)}
\end{align}

When the coupling to the cytoskeleton is switched on  ($\lambda>0$), we find that to linear order in $k$, Eq.~(\ref{eq:s(k)}) gives the positive solution $s=\frac{D_b k^{\text{off}}}{k^{\text{on}}}|k|$, which indicates that the homogeneous solution is unstable and spontaneous polarization occurs. 
The full solution $s(k)$ 
is plotted in Fig.~\ref{fig:s-k}, revealing a maximum at finite $k=k_{\text{max}}$ which corresponds to an instability of finite characteristic length $2\pi/k_{\text{max}}$. Fig.~\ref{fig:kmax-lambda}a shows that the most unstable mode $k_{\text{max}}$
 increases with $\lambda$, and therefore with the strength of the coupling $\alpha$ and the number of markers $M$. In the case of a cell of perimeter $L$, we conclude that  for a weak coupling  $2\pi/k_{\text{max}}>L$  and no patches will be seen. Increasing the coupling $\alpha$ (or $M$),  $\lambda$ reaches a series of thresholds defined by $2\pi/k_{\text{max}}=L/n_p$ and above which $n_p$ patches on the cell will grow. This generic behavior predicted in the case of cortical actin mediated transport agrees qualitatively with available experimental observations of spontaneous polarization in yeast. Indeed, it is found in \cite{Piel2008} that spontaneous polarization occurs  only above a threshold value of uniform concentration of pheromone (1nM $\alpha$-factor, which is assumed to drive the coupling $\alpha$ of our model), while it is shown in   \cite{Wedlich-Soldner2003}  that the number of patches indeed grows with the quantity of markers (Cdc42). Note that our model also predicts that increasing the size of the cell would increase the number of patches, which could be tested experimentally. Finally, the variation in the timescale for the polarization of a cell, $\tau=1/s$, with the parameter $\lambda$ is shown in Fig.~\ref{fig:kmax-lambda}b and corresponds qualitatively well to the data from \cite{Piel2008} showing decreasing time for spontaneous yeast polarization with increasing concentrations of pheromone.

\paragraph{Driven polarization.}
We now show that our model also accounts quantitatively for  driven polarization, for instance in the presence of an activator gradient. We assume that such  a gradient results in an asymmetric activation of the  markers on the membrane and therefore in a spatial inhomogeneity in the coupling parameter $\alpha$. We consider a perturbation $\alpha=\alpha_0+\alpha_k e^{ikx}$  (where $k=2\pi/L$ mimics a cell in a constant gradient) and therefore look for solutions at linear order $\mu(x)=\mu^0+\mu_ke^{ikx}$, $c(x,z)=c^0(z)+c_k(z)e^{ikx}$. 

\begin{figure}[hbt]
\begin{centering}
\includegraphics[width=0.5\textwidth]{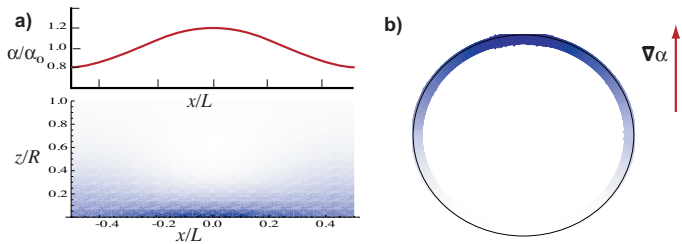}
\caption{\label{fig:density} Density of polarity markers $c(x,z)$ for driven polarization with $\alpha(x)$ (a) plotted in the $\{x,z\}$ model plane with parameters as in Fig.~\ref{fig:s-k} and $\lambda=6$, $k=1$, $M=1000$, $\alpha_k/\alpha_0=0.2$ (b) mapped onto a cartoon circular cell. Darker shading represents higher density.}
\end{centering}
\end{figure}

In case (a) of a  microtubule system,  Eq. (\ref{eq:cdyn}) with this
perturbation is solved by $c_k(z)=A_d e^{-\lambda_d z}+(c_k(0)-A_d)e^{-\rho_d z}$ with $\rho_d=\frac{\lambda_d}{2}+
\h(\lambda_d^2+4k^2)^{1/2}$,  where $\lambda_d=\alpha_0\mu^0/D_b$ and  $A_d$ is fixed by conservation of flux at the membrane.  This solution gives the polarized out of equilibrium steady state of the cell in response to an external gradient (see Fig. \ref{fig:density}), and qualitatively reproduces experiments on neuron growth cones \cite{Bouzigues2007}. Note that such a response is linear in the perturbation: it therefore occurs for arbitrary small activator gradients and no threshold is involved. Furthermore, the response is linearly stable, which implies that fluctuations in the external activator concentration will be damped out in the response, and the cell will polarize on average only along the activator gradient.

With the same method we can find the stationary state also in case (b) of a cortical actin system. This analysis however only applies in the regime of small coupling $\alpha_0$ were the steady state solution is stable as discussed earlier. 
In the unstable regime in an activator gradient the cell is likely to polarize spontaneously in the `wrong' direction due to concentration fluctuations, with a probability that increases for a short   polarization time $\tau$ (see Fig.\ref{fig:kmax-lambda}). There is therefore a trade off for the biological system between the abilities to polarize fast and to robustly polarize in the direction of a weak gradient. 
Sensitivity is compromised in cases such as mating yeast where it is important to polarize fast to gain a mating partner.
We expect this to be the case for many short range cell-cell interactions where local gradients are strong, masking the effects of fluctuations thereby decreasing the chance of spontaneous polarization in the wrong direction. On the other hand in cases such as neuron growth cones, the speed of polarization is sacrificed in favor of an accurate response to weak gradients.

To conclude, we have shown that the positive feedback triggered by the coupled dynamics of molecular polarity markers and the cytoskeleton is not sufficient to produce spontaneous polarization. We find that 
whether the system can polarize spontaneously or not actually depends crucially on the geometry of cytoskeletal  filaments.  
Active transport of markers along filaments oriented in a centered aster, such as microtubules, leads to
a stable homogeneous state, and polarized states occur only in response to a spatial gradient of activator concentration. 
This result is compatible with \cite{Altschuler2008}, where it was argued on the basis of a generic one dimensional model with local positive feedback only that homogeneous states are indeed stable,  and that polarized states occur only transiently due to stochastic fluctuations. On the contrary, in the case of filament asters centered on the membrane, such as cortical actin, we find that active  transport induces a non local coupling which  destabilizes the homogeneous states and therefore  leads to spontaneous polarization above a threshold uniform concentration of activator.  
Since the robustness of polarization along a weak gradient is compromised if the cell is able to polarize spontaneously this is of great importance in biological situations in which either robust response to a gradient or fast spontaneous polarization is (dis)advantageous.


\end{document}